# Distributed Denial of Service Prevention Techniques

B. B. Gupta, Student Member, IEEE, R. C. Joshi, and Manoj Misra, Member, IEEE

*Abstract—* The significance of the DDoS problem and the increased occurrence, sophistication and strength of attacks has led to the dawn of numerous prevention mechanisms. Each proposed prevention mechanism has some unique advantages and disadvantages over the others. In this paper, we present a classification of available mechanisms that are proposed in literature on preventing Internet services from possible DDoS attacks and discuss the strengths and weaknesses of each mechanism. This provides better understanding of the problem and enables a security administrator to effectively equip his arsenal with proper prevention mechanisms for fighting against DDoS threat.

## I. INTRODUCTION

A revolution came into the world of computer and communication with the advent of Internet. Today, Internet has become increasingly important to current society. It is changing our way of communication, business mode, and even everyday life [1]. Almost all the traditional services such as banking, power, medicine, education and defense are extended to Internet now. The impact of Internet on society can be seen from the fig. 1 which shows exponential increase in number of hosts interconnected through Internet [2]. Internet usage is growing at an exponential rate as organizations, governments and citizens continue to increase their reliance on this technology.

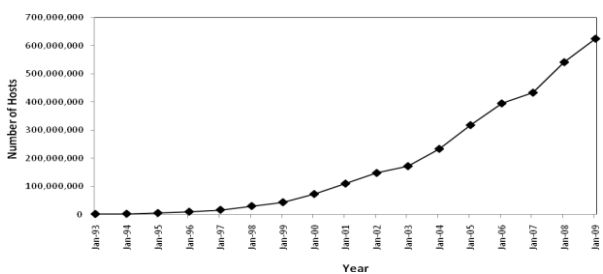

**Fig. 1** Internet Domain Survey Host Count

Unfortunately with an increase in number of host, count of attacks on Internet has also increased incredibly fast. According to [3], a mere 171 vulnerabilities were reported in 1995, which boomed to 7236 in 2007. Already, the number for the same for merely the third quarter of 2008 has gone up to 6058. Apart from these, a large number of vulnerabilities go unreported every year. In particular, today DoS attack is one of the most common and major threat to the Internet. In DoS attack, goal of the attacker is to tie up chosen key resources at the victim, usually by sending a high volume of seemingly legitimate traffic requesting some services from the victim. It reveals big loopholes not only in specific applications, but also in the entire TCP/IP protocol suite. DoS attack is considered to take place only when access to a computer or network resource is intentionally blocked or degraded as a result of malicious action taken by another user [4].

A DDoS attacker uses many machines to launch a coordinated DOS attack against one or more targets [5]. It is launched indirectly through many compromised computing systems by sending a stream of useless aggregate traffic meant to explode victim resources. As a side effect, they frequently create network congestion on the way from a source to the target, thus disrupting normal Internet operation. The number of DDoS attack has been alarmingly increasing for the last few years [6]. Many of today's DDoS attacks are carried out by organized criminals targeting financial institutions, e-commerce, gambling sites etc [7].

A classification of a wide range of DDoS attacks found in the wild is presented in [4, 8] that Internet providers and users need to be aware of. Usually, it can be launched in two forms [9]. The first form is to exploit software vulnerabilities of a target by sending malformed packets and crash the system. The second form is to use massive volumes of legitimate looking but garbled packets to clogs up computational or communication resources on the target machine so that it cannot serve its legitimates users. The resources consumed by attacks include network bandwidth, disk space, CPU time, data structures, network connections, etc. While it is possible to protect the first form of attack by patching known vulnerabilities, the second form of attack cannot be so easily prevented. The targets can be attacked simply because they are connected to the public Internet.

The first publicly reported DDoS attacks appeared in the late 1999 against a university [10]. These attacks quickly became increasingly popular as communities of crackers developed and released extremely sophisticated, user friendly and automated toolkits [11, 12, 13, 14, 15, 16, 17, 18, 19] to carry them out. At present, even people with little knowledge can use them to carry out DDoS attacks. The impact of DoS attacks can vary from minor inconvenience to users of a website, to serious financial losses for companies that rely on their on-line availability to do business.

Manuscript received April 1, 2009. This work was supported in part by the Ministry of Human Resource Development, Government of India.
B. B. Gupta is with the Department of Electronics & Computer Engg., Indian Institute of Technology, Roorkee, 247667 India. (phone: +91-9927713132; e-mail: bbgupta@ieee.org).
R. C. Joshi is with the Department of Electronics & Computer Engg., Indian Institute of Technology, Roorkee, 247667 India.
Manoj Misra is with the Department of Electronics & Computer Engg., Indian Institute of Technology, Roorkee, 247667 India.

This paper presents overview of DDoS problem, available DDoS attack tools, defense challenges and principles and a classification of available mechanisms that are proposed in literature on preventing Internet services from possible DDoS attacks and discuss the strengths and weaknesses of each mechanism. A summery of pending concerns draw attention to core problems in existing mechanisms.

The remainder of the paper is organized as follows. Section II contains overview of DDoS problem. Section III describes variety of available DDoS attack tools in the details. Section IV discusses defense challenges and principles. Classification of available DDoS prevention mechanisms is described in section V. Finally, Section VI concludes the paper and presents further research scope.

## II. DDoS Overview

A Distributed Denial of Service attack is commonly characterized as an event in which a legitimate user or organization is deprived of certain services, like web, email or network connectivity, that they would normally expect to have. DDoS is basically a resource overloading problem. The resource can be bandwidth, memory, CPU cycles, file descriptors, buffers etc. The attackers bombard scare resource either by flood of packets or a single logic packet which can activate a series of processes to exhaust the limited resource [20]. In the Fig. 2 simplified Distributed DoS attack scenario is illustrated. The figure shows that attacker uses three zombie's to generate high volume of malicious traffic to flood the victim over the Internet thus rendering legitimate user unable to access the service.

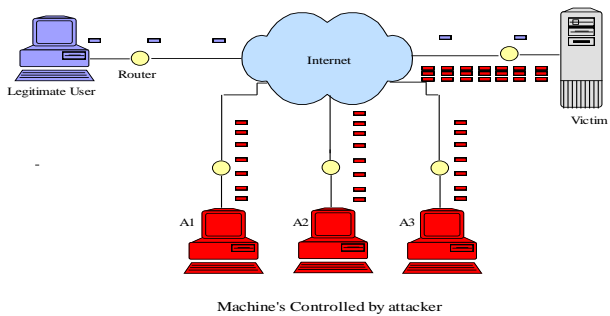

**Fig. 2** Illustration of the DDoS attack scenario

Extremely sophisticated, user friendly, automated and powerful DDoS toolkits are available for attacking any victim, so expertise is not necessarily required that attract naive users to perform DDoS attacks.

Although DoS attacking strategies differ in time, studies show that attackers mainly target the following resources to cause damage on victim [8, 21].

**Network bandwidth resources:** This is related with the capacity of the network links connecting servers to the wider Internet or connectivity between the clients and their Internet Service Providers (ISP). Most of the time, the bandwidth of client's internal network is less than its connectivity with the external network. Thus the traffic that comes from the Internet to the client may consume the entire bandwidth of the client's network. As a result, a legitimate request will not be able to get service from the targeted network. In a DoS attack, the vast majority of traffic directed at the target network is malicious; generated either directly or indirectly by an attacker. These attacks prevented 13,000 Bank of America ATM from providing withdrawn services and paralyzed such large ISPs as Freetel, SK Telecom, and KoreaTelecom on January 25, 2003.

1) **System memory resources:** An attack targeting system memory resources typically aims to crash its network handling software rather than consuming bandwidth with large volume of traffic. Specific packets are sent to confuse the operating system or other resources of the victim's machine. These include temporary buffer used to store arriving packets, tables of open connections, and similar memory data structures. Another system resource attack uses packets whose structures trigger a bug in the network software, overloading the target machine or disabling its communication mechanism or making a host crash, freeze or reboot which means the system can no longer communicate over the network until the software is reloaded.

2) **System CPU resources/ Computational Capacity:** An attack targeting system's CPU resources typically aims to employ a sequence of queries to execute complex commands and then overwhelmed the CPU. The Internet key Exchange protocol (IKE) is the current IETF standard for key establishment and SA parameter negotiation of IPsec. However, IKE's aggregate mode is still very susceptible to DoS attacks against both computational and memory resources because the server has to create states for SA and compute Diffie-Hellman exponential generation [22].

## III. DDoS Attacks Tools

One of the major reason that make the DDoS attacks wide spread and easy in the Internet is the availability of attacking tools and the powerfulness of these tools to generate attacking traffic. There are a variety of different DDoS attack tools on the Internet that allow attackers to execute attacks on the target system. Some of the most common tools are discussed below:

1) *Trinoo* [8, 11] can be used to launch a coordinated UDP flooding attack against target system. Trinoo deploys master/slave architecture and attacker controls a number of Trinoo master machines. Communication between attacker and master and between master and slave is performed through TCP and UDP protocol, respectively. Both master and slaves are password protected to prevent them from being taken over by another attacker. Wintrinoo

| DDoS attack tool | Commands used | Types of Attacks Generated | Communication Methods |
|---|---|---|---|
| Trinoo | Not encrypted | UDP flooding | Attacker to master-unencrypted TCP<br>Master to slave- unencrypted UDP<br>Slave to master - unencrypted UDP |
| TFN | Numeric code and not encrypted | ICMP flooding<br>TCP flooding<br>UDP flooding<br>Smurf | Attacker to master -required third-party program<br>Master to slave- unencrypted ICMP<br>Slave to master - none |
| TFN2K | Encrypted | ICMP flooding<br>TCP flooding<br>UDP flooding<br>Smurf<br>Mix flood | Master to slave- can be mixture of encrypted TCP, UDP and ICMP<br>Slave to master - none |
| Stacheldraht | Encrypted | ICMP flooding<br>TCP flooding<br>UDP flooding<br>Smurf | Attacker to master -encrypted TCP<br>Master to slave- TCP and ICMP<br>Slave to master - none |
| Shaft | Not encrypted | ICMP flooding<br>TCP flooding<br>UDP flooding<br>Mix flood | Attacker to master -unencrypted TCP<br>Master to slave- unencrypted UDP |
| Mstream | Not encrypted | TCP flooding | Attacker to master -unencrypted TCP<br>Master to slave- unencrypted UDP<br>Slave to master - unencrypted UDP |
| Knight | Not encrypted | TCP flooding<br>UDP Flooding<br>an urgent pointer flooder | Uses IRC as it's communication method |
| Trinity | Not encrypted | TCP flooding<br>UDP flooding | Uses IRC as it's communication method |

**Table I** Summery of DDoS attack Tools

is a Windows version of trinoo that was first reported to CERT on February 16th 2000.

2) *TFN* [12] uses a command line interface to communicate between the attacker and the control master program but offers no encryption between attacker and masters or between masters and slaves. Communication between the control masters and slaves is done via ICMP echo reply packets. It can implement Smurf, SYN Flood, UDP Flood, and ICMP Flood attacks.

3) *TFN2K* [13] is a more advanced version of the primitive TFN network. It uses TCP, UDP, ICMP or all three to communicate between the control master program and the slave machines. TFN2K can implement Smurf, SYN, UDP, and ICMP Flood attacks. Communication between the real attacker and control master is encrypted using a key-based CAST-256 algorithm. In addition to flooding, TFN2K can also perform some vulnerability attacks by sending malformed or invalid packets.

4) *Stacheldraht* [14] combines best features of both Trinoo and TFN. It also has the ability to perform updates on the slave machines automatically. It uses an encrypted TCP connection for communication between the attacker and master control program. Communication between the master control program and attack daemons is conducted using TCP and ICMP. Stacheldraht can implement Smurf, SYN Flood, UDP Flood, and ICMP Flood attacks.

5) *Shaft* [15] has been modeled on Trinoo network. Other than the port numbers being used for communication

purpose, working of it is very similar to the Trinoo. Thus, distinctive feature of Shaft is the ability to switch control master servers and ports in real time, hence making detection by intrusion detection tools difficult. Communication between the control masters and slave machines is achieved using UDP packets. The control masters and the attacker communicate via a simple TCP telnet connection. Shaft can implement UDP, ICMP, and TCP flooding attack.

6) *Mstream* [16] is more primitive than any of the other DDoS tools. It attacks target machine with a TCP ACK flood. Communication is not encrypted and is performed through TCP and UDP packets and the master connects via telnet to zombie. Masters can be controlled remotely by one or more attackers using a password protected interactive login. Source addresses in attack packets are spoofed at random. Unlike other DDoS tools, here, masters are informed of access, successful or not, by competing parties.

7) *Knight* [17] uses IRC as a control channel. It has been reported that the tool is commonly being installed on machines that were previously compromised by the BackOrifice Trojan horse program. Knight can implement SYN attacks, UDP Flood attacks, and an urgent pointer flooder [19]. It is designed to run on Windows operating systems and has features such as an automatic updater via http or ftp, a checksum generator and more.

8) *Trinity* [18, 19] is also IRC based DDoS attack tool. It can implement UDP, IP fragment, TCP SYN, TCP RST, TCP ACK, and other flooding attacks. Each trinity compromise machine joins a specified IRC channel and waits for commands. Use of legitimate IRC service for communication between attacker and agents eliminates the need for a master machine and elevates the level of the threat [4].

Table I shows a summary of different attack tools properties. Source code of these attack tools can be easily downloaded from the Internet. Even though these attack tools differ in the commands used, types of attacks used, communication techniques, and the presence of backdoors or self-upgrade capability, all share the common object of attempting to overwhelm a victim with an abundant amount of traffic that is difficult to detect or filter.

## IV. DEFENSE CHALLENGES AND PRINCIPLES

Launching DDoS attacks on the victim machine is only a matter of few keystrokes for the attacker. The victim can prevent from these attacks at its network boundary by configuring some sort of traditional security tools like access list [23], firewall [24, 25], or intrusion detection system [26, 27] at its end. But the regular benign traffic to the victim's network is not protected and moreover the victim cannot have access to other networks (e.g. the Internet).

With the present technology, many challenges are involved in designing and implementing an effective DDoS defense mechanism. Some of them are as follows [28]:

(a) Large number of unwitting participants, (b) No common characteristics of DDoS streams, (c) Use of legitimate traffic models by attackers, (d) No administrative domain cooperation, (e) Automated tools, (f) Hidden identity of participants, (g) Persistent security holes on the Internet, (h) Lack of attack information and (i) Absence of standardized evaluation and testing approaches.

Thus following five principles [29] are recommended by robinson et al. in order to build an effective solution:

9) Since, DDoS is a distributed attack and because of high volume and rate of attack packets, distributed instead of centralized defense is the first principle of DDoS defense.
10) Secondly, High Normal Packet Survival Ratio (NPSR) hence less collateral damage is the prime requirement for a DDoS defense.
11) Third, a DDoS defense method should provide secure communication for control messages in terms of confidentiality, authentication of sources, integrity and freshness of exchanged messages between defense nodes.
12) Fourth, as there is no centralized control for autonomous systems (AS) in Internet, a partially and incrementally deployable defense model which does not need centralized control will be successful.
13) Fifth, a defense system must take into account future compatibility issues such as interfacing with other systems and negotiating different defense policies.

Similarly, Tupakula et. al. [30] presented following characteristics that an ideal effective model against DDoS attacks should have:

14) It should be invoked only during the attack times and at other times it must allow the system to work normally. So it should readily integrate with existing architecture with minimum modifications.
15) It must provide simple, easy and effective solution to counteract the attacking sources in preventing the attack.
16) It should identify the attack at the victim and prevent the attack near to the attacking source.
17) It should prevent only the attack traffic from reaching victim. That is, the model should be able to differentiate a malicious traffic flow from a regular benign flow by incorporating different attack signatures for different attacking sources.
18) It should have fast response time and should respond quickly to any changes in attack traffic pattern.
19) It should provide mechanisms for retaining the attack evidence for any future legal proceedings.

## V. CLASSIFICATION OF DDOS PREVENTION MECHANISMS

Attack prevention methods try to stop all well known signature based and broadcast based DDoS attacks from being launched in the first place or edge routers, keeps all the machines over Internet up to date with patches and fix security holes. Attack prevention schemes are not enough to stop DDoS attacks because there are always vulnerable to novel and mixed attack types for which signatures and patches are not exist in the database.

Techniques for preventing against DDoS can be broadly divided into two categories: (i) General techniques, which are some common preventive measures [31] i.e. system protection, replication of resources etc. that individual servers and ISPs should follow so they do not become part of DDoS attack process. (iii) Filtering techniques, which include ingress filtering, egress filtering, router based packet filtering, history based IP filtering, SAVE protocol etc.

### A. General Techniques

1) Disabling unused services

The less there are applications and open ports in hosts, the less there are chance to exploit vulnerabilities by attackers. Therefore, if network services are not needed or unused, the services should be disabled to prevent attacks, e.g. UDP echo, character generation services [31].

2) Install latest security patches

Today, many DDoS attacks exploit vulnerabilities in target system. So removing known security holes by installing all relevant latest security patches prevents re-exploitation of vulnerabilities in the target system [31].

3) Disabling IP broadcast

Defense against attacks that use intermediate broadcasting nodes e.g. ICMP flood attacks, Smurf attacks etc. will be successful only if host computers and all the neighboring networks disable IP broadcast [32].

4) Firewalls

Firewalls can effectively prevent users from launching simple flooding type attacks from machines behind the firewall. Firewalls have simple rules such as to allow or deny protocols, ports or IP addresses. But some complex attack e.g. if there is an attack on port 80 (web service), firewalls cannot prevent that attack because they cannot distinguish good traffic from DoS attack traffic [24, 25].

5) Global defense infrastructure

A global deployable defense infrastructure can prevent from many DDoS attacks by installing filtering rules in the most important routers of the Internet. As Internet is administered by various autonomous systems according their own local security policies, such type of global defense architecture is possible only in theory [31].

6) IP hopping

DDoS attacks can be prevented by changing location or IP address of the active server proactively within a pool of homogeneous servers or with a pre-specified set of IP address ranges [31]. The victim computer's IP address is invalidated by changing it with a new one. Once the IP addresses change is completed all internet routers will be informed and edge routers will drop the attacking packets. Although this action leaves the computer vulnerable because the attacker can launch the attack at the new IP address, this option is practical for DDoS attacks that are based on IP addresses. On the other hand, attackers can make this technique useless by adding a domain name service tracing function to the DDoS attack tools.

### B. Filtering Techniques

1) Ingress/Egress filtering

Ingress Filtering, proposed by Ferguson et al. [33], is a restrictive mechanism to drop traffic with IP addresses that do not match a domain prefix connected to the ingress router. Egress filtering is an outbound filter, which ensures that only assigned or allocated IP address space leaves the network. A key requirement for ingress or egress filtering is knowledge of the expected IP addresses at a particular port. For some networks with complicated topologies, it is not easy to obtain this knowledge.

One technique known as reverse path filtering [34] can help to build this knowledge. This technique works as follows. Generally, a router always knows which networks are reachable via any of its interfaces. By looking up source addresses of the incoming traffic, it is possible to check whether the return path to that address would flow out the same interface as the packet arrived upon. If they do, these packets are allowed. Otherwise, they are dropped.

Unfortunately, this technique cannot operate effectively in real networks where asymmetric Internet routes are not uncommon. More importantly, both ingress and egress filtering can be applied not only to IP addresses, but also protocol type, port number, or any other criteria of importance. Both ingress and egress filtering provide some opportunities to throttle the attack power of DoS attacks. However, it is difficult to deploy ingress/egress filtering universally. If the attacker carefully chooses a network without ingress/egress filtering to launch a spoofed DoS attack, the attack can go undetected. Moreover, if an attack spoofs IP addresses from within the subnet, the attack can go undetected as well. Nowadays DDoS attacks do not need to use source address spoofing to be effective. By exploiting a large number of compromised hosts, attackers do not need to use spoofing to take advantage of protocol vulnerabilities or to hide their locations. For example, each legitimate HTTP Web page request from 10,000 compromised hosts can bypass any ingress/egress filtering, but in combination they can constitute a powerful attack. Hence, ingress and egress filtering are ineffective to stop DDoS attacks.

2) Router based packet filtering

Route based filtering, proposed by Park and Lee [35], extends ingress filtering and uses the route information to filter out spoofed IP packets. It is based on the principle that for each link in the core of the Internet, there is only a limited set of source addresses from which traffic on the link could have originated.

If an unexpected source address appears in an IP packet on a link, then it is assumed that the source address has been spoofed, and hence the packet can be filtered. RPF uses information about the BGP routing topology to filter traffic with spoofed source addresses. Simulation results show that a significant fraction of spoofed IP addresses can be filtered if RPF is implemented in at least 18% of ASs in the Internet. However, there are several limitations of this scheme. The first limitation relates to the implementation of RPF in practice. Given that the Internet contains more than 10,000 ASs, RPF would need to be implemented in at least 1800 ASs in order to be effective, which is an onerous task to accomplish. The second limitation is that RPF may drop legitimate packets if there has recently been a route change. The third potential limitation is that RPF relies on valid BGP messages to configure the filter. If an attacker can hijack a BGP session and disseminate bogus BGP messages, then it is possible to mislead border routers to update filtering rules in favor of the attacker. RPF is effective against randomly spoofed DoS attacks. However, the filtering granularity of RPF is low. This means that the attack traffic can still bypass the RPF filters by carefully choosing the range of IP addresses to spoof. Hence, RPF is ineffective against DDoS attacks. The router-based packet filter is vulnerable to asymmetrical and dynamic Internet routing as it does not provide a scheme to update the routing information.

3) History based IP filtering

Generally, the set of source IP addresses that is seen during normal operation tends to remain stable. In contrast, during DoS attacks, most of the source IP addresses have not been seen before. Peng et al. relies on the above idea and use IP address database (IAD) to keep frequent source IP addresses. During an attack, if the source address of a packet is not in IAD, the packet is dropped. Hash based/Bloom filter techniques are used for fast searching of IP in IAD. This scheme is robust, and does not need the cooperation of the whole Internet community [36].

However, history based packet filtering scheme is ineffective when the attacks come from real IP addresses. In addition, it requires an offline database to keep track of IP addresses. Therefore, Cost of storage and information sharing is very high.

4) Capability based method

Capability based mechanisms provides destination a way to control the traffic directed towards itself. In this approach, source first sends request packets to its destination. Router marks (pre-capabilities) are added to request packet while passing through the router. The destination may or may not grant permission to the source to send. If permission is granted then destination returns the capabilities, if not then it does not supply the capabilities in the returned packet. The data packets carrying the capabilities are then send to the destination via router. The main advantage achieved in this architecture is that the destination can now control the traffic according to its own policy, thereby reducing the chances of DDoS attack, as packets without capabilities are treated as legacy and might get dropped at the router when congestion happens [37].

Table II Summary of filtering techniques for DDoS attacks prevention

| Filtering Technique | Benefits | Limitations |
|---|---|---|
| Ingress/ Egress | -Prevents IP Spoofing | -Need global development<br>- Attacks with real IP addresses can not be prevented |
| RPF ( Route based Packet Filtering) | -Work well with static routing | -Problem when dynamic routing is used<br>-Need wide implementation to be effective |
| History based | -Does not require cooperation of whole Internet Community.<br>-Gives priority to the frequent packets in case of congestion or attack | - Ineffective when the attacks come from real IP addresses<br>- Requires an offline database to keep track of IP addresses<br>-Depend on information collected |
| Capability based | -Provides destination a way to control the traffic it desires<br>-Incremental deployment | -Attacks against the request packets can not prevented (e.g. ROC attack)<br>-High computational complexity and space requirement |
| SOS | -Works well for communication of predefined source nodes | -Solution has limited scope e.g. not applicable to web servers<br>-Require introduction of a new routing protocol that itself another security issue |
| SAVE | -Filtering improperly addressed packets is worthwhile<br>-incremental deployment | -During the transient period valid packets can be dropped |

However, these systems offer strong protection for established network flows, but responsible to generate a new attack type known as DOC (Denial of Capability), which prevents new capability-setup packets from reaching the destination, limits the value of these systems. In addition, these systems have high computational complexity and space requirement.

5) Secure overlay Service (SOS)

Secure Overlay Service proposed by Keromytis et al. [38] defines an architecture called secure overlay service (SOS) to secure the communication between the confirmed users and the victim. All the traffic from a source point is verified by a secure overlay access point (SOAP). Authenticated traffic will be routed to a special overlay node called a beacon in an anonymous manner by consistent hash mapping. The beacon then forwards traffic to another special overlay node called a secret servlet for further authentication, and the secret servlet forwards verified traffic to the victim. The identity of the secret servlet is revealed to the beacon via a secure protocol, and remains a secret to the attacker. Finally, only traffic forwarded by the secret servlet chosen by the victim can pass its perimetric routers.

Secure Overlay Service (SOS) addresses the problem of how to guarantee the communication between legitimate users and a victim during DoS attacks. SOS can greatly reduce the likelihood of a successful attack. The power of SOS is based on the number and distribution level of SOAPs. However, wide deployment of SOAPs is a difficult DoS defense challenge. Moreover, the power of SOS is also based on the anonymous routing protocol within the overlay nodes. Unfortunately, the introduction of a new routing protocol is in itself another security issue. If an attacker is able to breach the security protection of some overlay node, then it can launch the attack from inside the overlay network. Moreover, if attackers can gain massive attack power, for example, via worm spread, all the SOAPs can be paralyzed, and the target's services will be disrupted.

6) SAVE: Source Address Validity Enforcement

Li et al. [39] have proposed a new protocol called the Source Address Validity Enforcement (SAVE) protocol, which enables routers to update the information of expected source IP addresses on each link and block any IP packet with an unexpected source IP address. The aim of the SAVE protocol is to provide routers with information about the range of source IP addresses that should be expected at each interface. Similarly to the existing routing protocols, SAVE constantly propagates messages containing valid source address information from the source location to all destinations. Hence, each router along the way is able to build an incoming table that associates each link of the router with a set of valid source address blocks. SAVE is a protocol that enables the router to filter packets with spoofed source addresses using incoming tables. It overcomes the asymmetries of Internet routing by updating the incoming tables on each router periodically.

However, SAVE needs to change the routing protocol, which will take a long time to accomplish. If SAVE is not universally deployed, attackers can always spoof the IP addresses within networks that do not implement SAVE. Moreover, even if SAVE were universally deployed, attackers could still launch DDoS attacks using non spoofed source addresses.

Table II summarizes filtering techniques for DDoS attacks prevention.

To conclude, attack prevention aims to solve IP spoofing, a fundamental weakness of the Internet. However, as attackers gain control of larger numbers of compromised computers, attackers can direct these "zombies" to attack using valid source addresses. Since the communication between attackers and "zombies" is encrypted, only "zombies" can be exposed instead of attackers. According to the Internet Architecture Working Group [40], the percentage of spoofed attacks is declining. Only four out of 1127 customer-impacting DDoS attacks on a large network used spoofed sources in 2004. Moreover, security awareness is still not enough, so expecting installation of security technologies and patches in large base of Internet seems to be an ambitious goal in near future. To add on, there exists no way out to enforce global deployment of a particular security mechanism. Therefore, relying on attack prevention schemes is not enough to stop DDoS attacks.

## VI. CONCLUSIONS AND FUTURE SCOPE

DoS attack causes either disruption or degradation on victim's shared resources, as a result preventing legitimate users from their access right on those resources. DoS attack may target on a specific component of computer, entire computer system, certain networking infrastructure, or even entire Internet infrastructure. Attacks can be either by exploits the natural weakness of a system, which is known as logical attacks or overloading the victim with high volume of traffic, which is called flooding attacks. A distributed form of DoS attack called DDoS attack, which is generated by many compromised machines to coordinately hit a victim. DDoS attacks are adversarial and constantly evolving. Once a particular kind of attack is successfully countered, a slight variation is designed that bypasses the defense and still performs an effective attack.

In this paper, we covered an overview of the DDoS problem, available DDoS attack tools, defense challenges and principles, and a classification of available DDoS prevention mechanisms. This provides better understanding of the problem and enables a security administrator to effectively equip his arsenal with proper prevention mechanisms for fighting against DDoS threat. The current prevention mechanisms reviewed in this paper are clearly far from adequate to protect Internet from DDoS attack. The main problem is that there are still many insecure machines over the Internet that can be compromised to launch large-scale coordinated DDoS attack. One promising direction is to develop a comprehensive solution that encompasses several defense activities to trap variety of DDoS attack. If one level of defense fails, the others still have the possibility to defend against attack. A successful intrusion requires all defense level to fail.

ACKNOWLEDGMENT

The authors gratefully acknowledge the financial support of the Ministry of Human Resource Development (MHRD), Government of India for partial work reported in the paper.

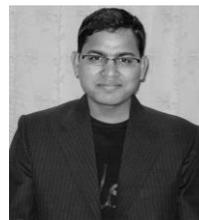

**B. B. Gupta** received the bachelor's degree in Information Technology in 2005 from Rajasthan University, India. He is currently a PhD student in the Department of Electronics and Computer Engineering at Indian Institute of Technology, Roorkee, India.

His research interests include defense mechanisms for thwarting Denial of Service attacks, Network security, Cryptography, Data mining and Data structure and Algorithms. He is a student member of IEEE.

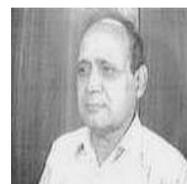

**R. C. Joshi** received the bachelor's degree in Electrical Engineering from Allahabad University, India in 1967. He received his master's and PhD degree in Electronics and Computer Engineering from University of Roorkee, India in 1970 and 1980, respectively. Currently, he is working as a Professor at Indian Institute of Technology Roorkee, India.

He has served as Head of the Department twice from Jan 1991 to Jan 1994 and from Jan 1997 to Dec 1999. He has been Head of Institute Computer Centre (ICC), IIT Roorkee from March 1994- Dec 2005. Prof. Joshi is in expert panel of various national committees like AICTE, DRDO and MIT. He has a vast teaching experience exceeding 38 years at graduate and postgraduate levels at IIT Roorkee. He has guided over 25 PhD thesis, 150 M.E./M.Tech dissertations and 200 B.E./B.Tech projects. Prof. Joshi has published over 250 research papers in National/International Journals/Conferences and presented many in Europe, USA and Australia. He has been awarded Gold Medal by Institute of Engineers for best paper. He has chaired many national and international conferences and workshops. Presently, he is actively involved in research in the field of Database management system, Data mining, Bioinformatics, Information security, Reconfigurable systems and Mobile computing.

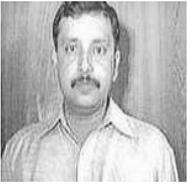

**Manoj Misra** received the bachelor's degree in Electrical Engineering in 1983 from HBTI Kanpur, India. He received his master's and PhD degree in Computer Engineering in 1986 and 1997 from University of Roorkee, India and Newcastle upon Tyne, UK, respectively. He is currently a Professor at Indian Institute of Technology Roorkee.

He has guided several PhD theses, M.E./M.Tech. Dissertations and completed various projects. His areas of interest include Mobile computing, Distributed computing and Performance Evaluation. He is a member of IEEE.